\documentclass[runningheads]{llncs}
\usepackage{graphicx}
\usepackage{bibentry}
\usepackage[hyphenbreaks]{breakurl}
\usepackage[hyphens]{url}

\begin{document}
\title{Reference Service Model for\\Federated Identity Management}
\titlerunning{FIMSM}
%
\author{Daniela Pöhn\and
Peter Hillmann}
\authorrunning{D. Pöhn and P. Hillmann}
%
\institute{Universität der Bundeswehr München, 85577 Neubiberg, Germany \\
\email{\{daniela.poehn,peter.hillmann\}@unibw.de}}
\maketitle              
\begin{abstract}
With the pandemic of COVID-19, people around the world increasingly work from home. Each natural person typically has several digital identities with different associated information. During the last years, various identity and access management approaches have gained attraction, helping for example to access other organizations’ services within trust boundaries. The resulting heterogeneity creates a high complexity to differentiate between these approaches and scenarios as participating entity; combining them is even harder. Last but not least, various actors have a different understanding or perspective of the terms, like “service”, in this context. Our paper describes a reference service with standard components in generic federated identity management. This is utilized with modern Enterprise Architecture using the framework ArchiMate. The proposed universal federated identity management service model (FIMSM) is applied to describe various federated identity management scenarios in a generic service-oriented way. The presented reference design is approved in multiple aspects and is easily applicable in numerous scenarios. 
\end{abstract}
\section{Introduction}

Due to the digitization, people are using an increasing number of identities in various fields. Every person has in average about 130 online accounts~\cite{11} for, e.\,g., work, e-government, online shops, and social networks. Each digital identity contains different user information, called attributes. Providers like Google allow users to securely reuse their accounts to authenticate at other services by utilizing the protocols Open Authorization (OAuth) / OpenID Connect (OIDC). The alternative Federated Identity Management (FIM) protocol Security Assertion Markup Language (SAML) is typically used in research and education. FIM has the advantage over traditional centralized identity and access management (I\&AM) that the attributes are stored authoritatively at the home organization, i.\,e., Identity Provider (IDP). Thereby, the data is consistent and the user has only one digital identity to remember. The Service Provider (SP) asks the IDP about the authentication respectively authorization. As a consequence, the SP securely reuses the information. If not all entities use the same protocol or a variant of it, additional effort is required. By shifting the focus from applications to the user, user-centric approaches were introduced, adding even more diversity to the landscape. According to Dhamija and Dusseault~\cite{6}, users follow the path of least resistance, while cognitive scalability is as important as technical scalability. With the different ecosystems in mind, these goals seem difficult to reach. The diversity of identities is growing owing to several factors, like smartphones and Internet of Things (IoT) devices. The multitude of standards and solutions hardens the task of companies due to the lack of interoperability between them. This makes it increasingly difficult to get an overview of these complex structures. An outline and an understanding of the provided services is though needed to improve the current identity management systems. Therefore, a systematic approach is required. 

To address these problems, a reference architecture for federated identity management service model (FIMSM) is designed with the supporting toolkit of Enterprise Architecture applying the framework ArchiMate as it is the most used modelling approach~\cite{4}. ArchiMate is an open enterprise architecture modelling language highlighting mainly the design and structure of a system. Furthermore, it supports the description, analysis, and visualization of architectures in a generic way. A reference architecture consists of reusable models and patterns, which are adaptable and customizable. It describes a particular recurring design problem that arises in specific contexts and presents a well-proven solution for the problem~\cite{4}. FIMSM offers overviews from different hierarchy levels and several detailed perspectives. The main advantage of a representation in the form of a model is that it is more precise and easier to understand than a textual description. In addition, the transfer is supported with model-driven architectures. Our reference architecture is conform to Information Technology Infrastructure Library (ITIL), The Open Group Architecture Framework (TOGAF) and Federated Mission Network as well as further frameworks. Organizations can refer to this template when adapting or constantly improving FIM.

This paper contributes the following improvements to the modelling and identity management landscape: With FIMSM, a reference architecture focusing on federated identity management is designed, which can be reused. With the model and patterns, the architecture describes elements, relations, functionality, and data flows required for FIM. It is the first concept for a universal identity management reference architecture covering all identity management variants. This overview is an important step towards this goal. With the reference model as background, it forms the basis to improve and enrich identity management in future work.

The paper is structured as follows: After the introduction, we discuss a motivating scenario with three entities and different protocols, which identifies several key components in Section~\ref{sec:scenario}. Related approaches are shown in Section~\ref{sec:sota}. Based on the key components and findings, we present the FIMSM in Section~\ref{sec:uimsm}. The discussion in Section~\ref{sec:discussion} is three-folded with scenarios, requirements, and an model adaption. Section~\ref{sec:conclusion} serves as conclusions and outlook.

\section{Motivating Scenario}
\label{sec:scenario}

The problem description arises from three entities, which want to cooperate: 1) IDP of a university, 2) IDP and SP of a commercial company, and 3) IDP and SP of a university hospital. All three entities use different identity management systems, which may result in several accounts for the end users.

1) University\\ 
\textbf{Local I\&AM:} In order to provide several services, like email, file transfer, and video conference systems, to students, employees, and professors alike, universities have a local I\&AM in place. This I\&AM is typically based on directory services, e.\,g., lightweight directory access protocol (LDAP). For Single Sign-On (SSO), another service is running, tracking the different sessions.\\
\textbf{FIM:} As a result of cooperation, national and international research and education federations, like DFN-AAI and SWITCHaai, were established. These provide services to each other within certain trust boundaries. The university is part of one of these national federations. The participating IDPs and SPs communicate with a pre-decided protocol, i.\,e., SAML. In order to make use of the federation, the SAML implementation Shibboleth is applied. The federation operators provide additional services, e.\,g., distribution of aggregated SAML metadata, via a trusted third party (TTP). The aggregated SAML metadata consists of information about the members of the federation enabling cooperation. Collaborations outside of the federation are difficult though. Some services already accept different protocols; otherwise proxies may be needed.

2) Commercial company\\ \textbf{Local I\&AM:} The company uses Active Directory (AD) as local I\&AM.\\
\textbf{FIM:} In order to access different web services including a private cloud, the component Federation Services (AD~FS) is deployed. Suppliers and project partners have access to certain shared folders within the federated private cloud via some kind of OIDC. For specific services, a second factor is required. As both partners do not use the same protocol, a translation service is needed. If users of the company want to access external services, a proxy has to be added, since AD~FS only works with one SP.\\
\textbf{Others:} Other systems, like Pluggable Authentication Modules (PAM), are in control of access to servers. The company has factories, optimizing the manufacturing by adding IoT sensors. The I\&AM does not only include identification and management of access to different types of devices, and, thereby, data. It also addresses interactions between devices, services offered, and how to identify the device as well as control the access to sensitive data. Since IoT devices often are installed, where everyone can tamper with the device, they should not have access to the whole network. This puts different constraints on the adoption of existing I\&AM frameworks. IoT management is so far another identity management silo for the company. Therefore, interoperability is needed.

3) University hospital\\ \textbf{Local I\&AM:} The hospital is associated with the university. At the same time, it runs different research labs, where external researchers can access medical devices from remote via a virtual private network. These medical devices still run on older operating systems, as updates need to undergo a certification process. Since the hospital only provides a local I\&AM, external researchers receive local accounts. 
This local identity management system has implications for employees, which own another account from the university. Therefore, many end user hold several usernames and credentials. With all the different accounts, the end user want to get a better overview. 

Based on this scenario, we obtain the following requirements, which a reference architecture should fulfil.
\begin{description}
\item[R1:] Reusable architecture for I\&AM with generic and universal terminology.
\item[R2:] Systematic overview and detailed perspectives on single aspects.
\item[R3:] Adaptability to different protocols and use cases.
\item[R4:] Dependencies between various providers with related interfaces including requirements regarding an appropriate service management.
\end{description}

We take these requirements into account when designing a reference architecture for FIM.
Furthermore, the following research questions will be addressed.
\begin{description}
\item[Q1:] How to describe I\&AM scenarios with a scenario-independent approach?
\item[Q2:] Which elements are required to fulfil the requirements? 
\item[Q3:] Is it possible to adapt the reference architecture to different areas?
\vspace{-0.4mm}
\end{description}

\section{Related Approaches}
\label{sec:sota}
In this section, we give an overview of related approaches for I\&AM and models of selected processes and components within. First, we outline related work in the area of I\&AM (A). Then, we discuss modelling approaches for trust (B), followed by modelling approaches for identity management (C).

A) According to Yuan Cao and Lin Yang~\cite{13}, three different identity management models exist: isolated, centralized, and federated identity management. The authors further classify I\&AM into the paradigms network-centric, service-centric, and user-centric, which evolved into the models isolated, centralized, federated, and user-centric. Isolated identity management, i.\,e., local accounts, is typically no longer used, although exceptions exist. Centralized I\&AM is the base for FIM, which mainly comprises of the protocols SAML, OAuth, and OIDC. SAML federations are rather static due to the federation structure and metadata exchange via a TTP. OAuth and OIDC are more dynamic, but not less complex. OIDC is the authentication layer on top of OAuth. Both protocols, OAuth and OIDC, are further improved within OpenID Foundation and Internet Engineering Task Force. Another model relates to user-centric approaches. While different privacy enhancing technologies never established long living approaches in I\&AM, two main directions currently exist. User Managed Access decouples identity resolution from the maintenance of identity information. It is built upon OAuth, although the principle can be applied to other protocols as well. User Managed Access is currently extended for the usage with self-sovereign identity. The principles of self-sovereign identities are stated by Toth and Anderson-Priddy~\cite{20}. They are typically implemented by blockchain, a form of decentralized identifiers. DecentID by Friebe et al.~\cite{7}, e.\,g., uses a public blockchain as trust anchor and utilizes external storage, i.\,e., distributed hash tables, for attributes due to scalability. The stated identity management models are adjusted for cloud identity management by Zwattendorfer et al.~\cite{22}.
An overview of numerous research approaches within the identity management models is provided by Pöhn and Hommel~\cite{16}. These approaches are compared with derived requirements, showing that no flexible and universal solution exist.

B) The need for a holistic architecture view is mandatory to avoid duplication in realization and to obtain synergy effects between the different systems~\cite{1}. Especially the collaboration of several institutions in a federation becomes a challenge due to the growing IT complexity and architecture as a whole. The diversity of over 60 frameworks~\cite{12} offers a wide range of modelling description possibilities. Organizations are required to present the information systems underlying their business in a transparent manner. The establishment of Enterprise Architecture is a mandatory discipline for companies to overcome this challenge. One of the newer modelling languages is ArchiMate~\cite{19}, published in the year 2004. The main advantage is the simplicity. Therefore, it fits to our intention to provide a detailed overview and the interfaces for a federated system. Additionally, a transformation to other modelling approaches is possible~\cite{9}.

C) The stated identity models are described in a high-level or use case respectively implementation specific architectures, but no generic model for identity management is known to the authors. The developers of ArchiMate themselves offer only a rudimentary approach to identity management~\cite{18}. This does not explore the technical possibilities. Although Dabrowski and Pacyna~\cite{5} claim a reference architecture for identity management, no official meta-model is used. Pöhn and Hommel~\cite{15} describe FIM and Inter-FIM with an older meta-model, not taking different protocols and elements into account. The approaches of Yang et al.~\cite{21} and Katsikogiannis et al.~\cite{10} focus on the process of authentication and authorization. However, they do not take a holistic view from an architecture enterprise perspective. Elements for reference architectures are described in frameworks and standards, like ISO/IEC 24760. These are neither universal nor up-to-date.
A closely related approach to the problem is presented in form of an Enterprise Architecture Pattern by Perroud and Inversini~\cite{14}. Their reference design of the pattern LetMeAccess provides different views on the centralized I\&AM architecture, following no official meta-model. Amaral et al.~\cite{2} model the aspect of trust in general with ArchiMate, providing guidelines for this approach. Further similar models based on ontologies exists as pattern in ArchiMate for resource and capabilities by Azevedo et al.~\cite{3} as well as services by Griffo et al.~\cite{8}.
Another approach by Petrovska et al.~\cite{17} describes security object relationships. The use case of oracle identity manager is not specific modelled according a systematic framework. Cioroaica et al.~\cite{23} provide a reference architecture for trust-based digital ecosystems. Their focus is on evaluation of trustworthiness on runtime. Even if it does not provide a direct input for FIM, we ensure the compatibility, especially regarding the interfaces.

\section{Federated Identity Management Service Model}
\label{sec:uimsm}

\begin{figure}
  \centering
    \includegraphics[width=1\linewidth]{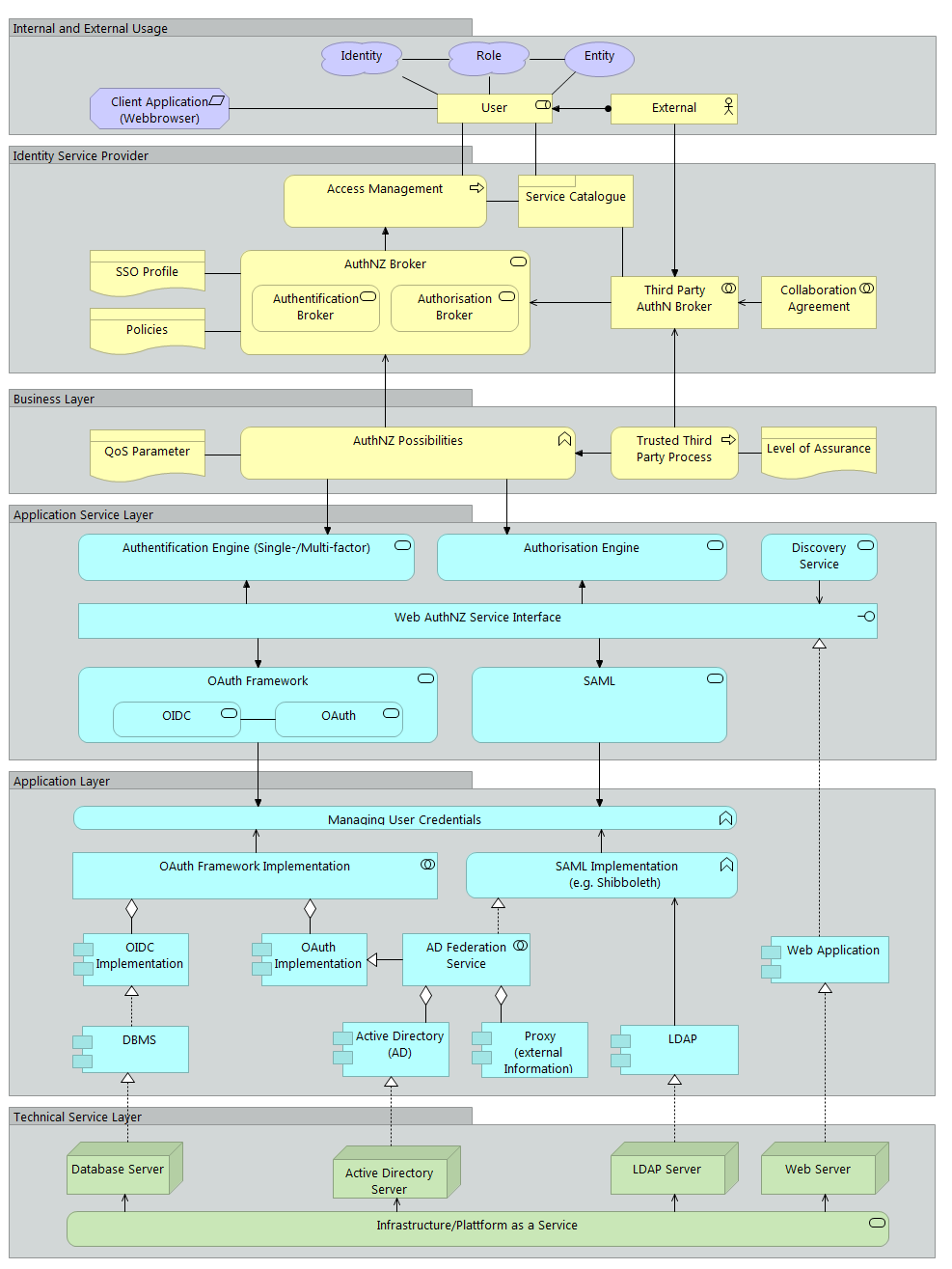}
    \caption{Architecture for Federated Identity Management}\label{fig:uimsm}
\end{figure}

In this section, we utilize ArchiMate to achieve FIMSM, a generic scenario-independent service model for FIM. The universal FIMSM needs to fulfil the requirements, stated in Section~\ref{sec:scenario}. It thereby has to be reusable with generic and universal terminology (R1), while presenting a systematic overview (R2). The overview should provide interfaces between different entities (R4), throughout being adaptable (R3). These requirements are addressed in the following model. According to ArchiMate's approach, Figure~\ref{fig:uimsm} shows the overview for the reference architecture FIMSM, using different layers to differentiating roles from business, applications, and technology. The reference architecture includes external layer, business service layer, business layer, application service layer, application layer, and the technical service layer. Consequently, a clear distinction is made between the provision of services and their implementation in the respective layer. We explain the universal approach according to top-down.
 
The \textbf{highest layer} describes the internal and external actors, i.\,e., the end users. 
On the top external / environment layer, a user wants to access a service. For authentication, he or she uses an I\&AM system, which is accessed via a client application, e.\,g., a web browser. The user’s IDP created an account with a digital identity for the user beforehand. Depending on the business of the IDP, it may perform an initial identity check using appropriate identification procedures, e.\,g., by verification of an ID card or bank account. Web services typically have no such or lower trusted verification methods. The roles are assigned to the account according to its position in their associated organization. Each user can take on different roles and assume different identities in relation to tasks. The user is either internal or external from a trusted organization. The latter describes typical FIM use cases. 

The \textbf{business service layer} shows the internal and external usage by the end users. This layer helps to separate the internal structure and organization from the external observable behaviour expressed at the service layer.
In the business service layer of FIMSM, i.\,e., the identity service provider layer, a corresponding service catalogue with trusted services is maintained. Such a list is usually not publicly available, but would be of great benefit. Access is provided by the process of access management, where the user can identify him- or herself by different methods. The most common authentication is via username and password. Increasingly, security is being enhanced by a second factor.
Access management process triggers the business service of the authentication and authorization (AuthNZ) broker offering different methods. The core business is extended by another broker, which takes over the handling for trustworthy third parties. For this purpose, appropriate collaboration agreements are negotiated beforehand. This can range from formal contracts to black and white lists. In the policies, the corresponding technical specifications for the respective services are kept available to define the transfer criteria between providers and customers. 

The \textbf{business layer} includes actors, roles, collaborations, and interfaces with regards of the provided service. These structural elements interact by processes, functions, interactions, events, and services. Combined they represent elements with relevance from a business perspective. For the sake of clarity and the goal of presenting the overall context of the architectural overview, some elements are not included and will appear in the adjusted detailed models during realization.
For FIMSM, the main service provided to end users and external entities is authentication and authorization, combined as AuthNZ possibilities. AuthNZ has quality of service (QoS) parameters, which the provider needs to fulfil. The service is only provided to externals, if both entities have a form of agreement or contract. The handling for external customers is embedded into the trusted third party process, which enables the trust establishment via an independent party. Participating entities may have requirements regarding security management, described by level of assurance. Level of assurance is the quantification of factors leading to confidence into an entity, relying on internal processes. This level of assurance can be stated within protocols.

The \textbf{application service layer} is built upon the application layer and shows accessible services and interfaces for other entities.
The main service, especially for service providers, is the web service, where users need to authenticate. The web service therefore requests the authentication and authorization engines. In order to do this, the home organization of the user needs to be discovered, described by the discovery service. Often, a user request is redirected to or personal data of the user is requested from the IDP. The discovery service assists the web AuthNZ service. The web service itself uses and triggers different protocols, e.\,g., representational state transfer (REST) interface. Within FIM, this is mainly SAML and the OAuth framework with OAuth for authorization and OIDC for authentication. As several libraries and implementations exist for both protocol variants, many web service offer both.

The \textbf{application layer} gives an overview of the application layer concepts and their relationships. The concept describes different application components, software, and interfaces. In order to provide services to other entities and end users, several software components need to be implemented. The web service requires a web application, running on a web server, described within the technical service layer. 
The core part is surrounding I\&AM itself, i.\,e., managing user credentials. Depending on the applied software, different protocols can be implemented. The user information itself is stored either in a database, AD, or LDAP. OIDC may be based on a database, but also uses other means of user storage. SAML typically utilizes LDAP. With AD in a federated use case, the extension AD~FS needs to be implemented. AD~FS provides SSO for different protocols, like OAuth and SAML. AD~FS typically interacts with one SP. If a proxy is added, it can serve several SPs. An IDP configures roles and thereby permissions at the I\&AM system. It is typical that services, independent of the location, have more fine-grained permissions locally. An entity is able to run all forms of FIM, but may decide on one specific software, implementation, and protocol. The most common variants are displayed. Further extensions can be added dependent on the use case.

The \textbf{technical service layer} consists the physical layer with actual hardware, including possible nodes, devices, infrastructure interfaces, communication paths, and networks.
At the FIMSM technical service layer level, separate servers are provided for the various applications. This separation results in a better illustration and takes the idea of micro-service architectures, e.\,g., based on Docker, into account. In addition, specific security requirements can be implemented in a targeted manner. The different servers are made available as infrastructure or platform as a service scalable according to the cloud business service model.

\section{Discussion}
\label{sec:discussion}
With the ever evolving extensions and revisions of FIM protocols, it is important to get the picture of the current status and gain a common understanding of the services offered. A reference architecture can help by providing a systematic way to model the domain services and associated interfaces. It though needs to fulfil several requirements to be of good use. The provided feedback from selected experts was embedded into the reference model. In FIMSM, we show the most common workflow. Several aspects of I\&AM can be separated into business processes and the formal specification of it. In the following, we apply the reference architecture to the three organizations described in the scenario. This is followed by analyses of the requirements and research questions. Last but not least, we adapt the reference architecture for Kerberos. It is a protocol for client authentication, which therefore belongs to centralized I\&AM for clients. Nevertheless, the principle of FIMSM can be adapted, showing that 1) it can be extended in future work for other identity management models as well as use cases and 2) it can help to identify combinations.

\subsection{Application of the Reference Architecture on the Scenario}
Since the goal of FIMSM is to be reusable for and adaptable to different use cases and protocols, the application of the reference architecture on the motivating scenario in Section~\ref{sec:sota} is shown. The university consists of a local I\&AM based on LDAP, which enables SSO. For FIM, the SAML implementation Shibboleth is used. The architecture of the FIM implementation Shibboleth is typically shown as external systems, like LDAP, and one big module. Hence, FIMSM is more fine-grained. As the university is part of federations, the federation operators use TTPs to distribute federation metadata. Therefore, the following can be seen.
\begin{description}
\item[External Layer:] The external layer consists of end users at the university, which have different identities and roles dependent on their function, like student, professor, or employee. These users utilize their browser to access services at the university and other entities.
\item[Business Service Layer:] In order to be able to access services, the user needs to authenticate. For this, a broker enables SSO.
\item[Business Layer:] The business layer of the university consists of the authentication and authorization functionality provided for internals and externals. With the trusted third party process, the university enables user to make use of FIM. The university signed a contract with the federation operator. After being an official member of the federation, the university has to follow certain procedures to ensure assurance, described by level of assurance. This may be “Advanced” within DFN-AAI. Even though the entity is a university, it has internal quality of service parameters for the service.
\item[Application Service Layer:] The application service level shows the discovery service used by SP. As the university offers services, they run their own local discovery service. The discovery service displays possible IDPs based on the SAML metadata file pulled from the federation operators TTP. The second factor may be required, dependent on the service providers’ needs.
\item[Application Layer:] The application layer of the university is also centred on user credential management. With Shibboleth as SAML implementation and LDAP along some web applications, the layer is rather lightweight.
\item[Technical Service Layer:] The university runs a LDAP server and web servers for identity management. Outside of the FIM scenario, the university has several other servers.
\end{description}

The identity management of the company is centred around AD with AD~FS and OAuth / OIDC. The latter is used to grant access to the private cloud for internal as well as external users. Additionally, PAM and IoT identity management are enabled. In this paper, we concentrate on FIM. AD~FS architecture is typically shown as AD~FS servers upon AD with several services made available through a proxy. Therefore, the following model is more fine-grained.
\begin{description}
\item[External Layer:] The external layer consists of end users, which are either associated with the company or business partners, having different identities with attributes, roles, and permissions.
\item[Business Service Layer:] The user wants to access services, e.\,g., the private cloud, and needs to authenticate before use. With AD~FS, SSO should be implemented. The company holds agreements with partners for FIM.
\item[Business Layer:] As the company has both roles, IDP and SP, the TTP process and the level of assurance need to be regarded in both directions.
\item[Application Service Layer:] With both roles, the company provides web service as well as authentication and authorization engine. Instead of SAML, the company makes use of OIDC build on OAuth.
\item[Application Layer:] According to the protocols and software, the application layer is more complex. The company applies FIM protocols build upon AD with AD~FS. Although OIDC often has an interface to a database, it can be added to other sources and implementations. Additionally, a web application, i.\,e., the private cloud, is shown.
\item[Technical Service Layer:] The technical service layer is a result from the layers above and consists of both, Active Directory Server and Web Server.
\end{description}

The hospital runs a local I\&AM, which is not in the focus of the FIMSM. Nevertheless, basic concepts can be applied. In context with the scenario, the reference architecture also shows the need for further software and services. Therefore, it can be utilized to identify and model them in a controlled process.
\begin{description}
\item[External Layer:] The users of the hospital are part of the external layer, having identities and roles, even though they are only used internally and locally.
\item[Business Service Layer:] Internally, authentication, authorization, and SSO are provided by a local I\&AM. Without FIM, collaboration aspects cannot be specified.
\item[Business Layer:] The hospital internally offers authentication and authorization possibilities, but as FIM software is missing, the users cannot interact with other services. Hence, no trusted third party process is available.
\item[Application Service Layer:] For internal services, which use LDAP, authentication and authorization engines are available.
\item[Application Layer:] The LDAP server runs a LDAP implementation, like OpenLDAP. This implementation builds the basis for different FIM software.
\item[Technical Service Layer:] In consequence, this layer consists of a LDAP server.
\end{description}

\subsection{Analysis of the FIMSM based on the Problem Statement}
Based on the motivating scenario, we obtained four requirements, which led to the research questions. In order to verify our results, we analyse the fulfilment of them.
\begin{description}
\item[R1:] Reusable architecture for I\&AM with generic and universal terminology. As shown in the previous section, the architecture can be reused for different protocols, since it provides a generic and universal terminology.
\item[R2:] Systematic overview and detailed perspectives on single aspects, like provided services. The application on the entities and services of the scenario was straight forward. Further division into entity roles could detail the layers.
\item[R3:] Adaptability to different protocols and use cases. The reference architecture is adaptable to SAML, OAuth, and OIDC with different implementations and constellations. Even though the focus is on FIM, it can be applied to centralized I\&AM, identifying missing services. In future work, we plan to adapt the approach for it.
\item[R4:] Dependencies between different providers with related interfaces including requirements regarding an appropriate service management. With the TTP process including level of assurance and collaboration agreement, a business process interface is modelled. The other dependency is a technical one with the discovery service, embedded into the web service. Since quality of service is relevant, additional interfaces, e.\,g., regarding federated security management can be modelled and added in the future.
\end{description}

The analysis shows that the requirements are fulfilled. Additionally, future work is identified, helping to build a truly universal service model for I\&AM. Regarding the stated research questions, the following is recognized.
\begin{description}
\item[Q1:] How to describe I\&AM scenarios with a scenario-independent approach? Different FIM scenarios can be described with FIMSM. Local I\&AM modelled on the approach shows deficits for FIM use cases. The systematic can be applied to other user cases, e.\,g., Kerberos, shown in the next section.
\item[Q2:] Which elements are required to fulfil the requirements? The architecture includes different protocol variants, their implementations and service aspects as well as process steps for federated use case in a generic, implementation-independent form. The layers help to identify and distinguish between applications, underlying infrastructure as well as services at different levels.
\item[Q3:] Is it possible to adapt the reference architecture to different areas? With regards to the hospital example, the reference architecture can be applied and adapted to other areas. We apply the architecture to Kerberos without changes in the systematic in the following section.
\end{description}    
    
\subsection{Adaptation for Kerberos}

\begin{figure}[t]
  \centering
    \includegraphics[width=0.48\textwidth]{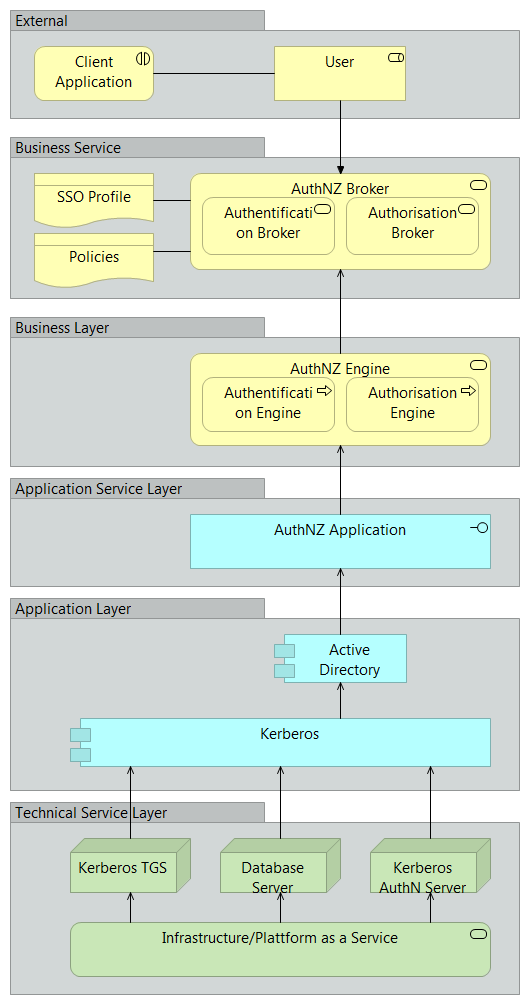}
    \caption{Adapted architecture for Kerberos}\label{fig:kerberos}
    \vspace{-1.0mm}
\end{figure}

Active Directory is built upon two main protocols: LDAP and Kerberos. Additionally, it needs Domain Name System (DNS). Kerberos is a computer network authentication protocol working with tickets to allow nodes communication over a non-secure network. It gives nodes the possibility to prove their identity to one another over this insecure network in a secure manner. The design is focused on a client-server model providing mutual authentication for both, the user and the server. Kerberos uses symmetric key cryptography and, therefore, requires a trusted party or public-key cryptography for the key exchange. The messages are hence protected against eavesdropping and replay attacks. Windows 2000 and later versions apply Kerberos as their default authentication method. It is therefore often found in companies, which have Windows clients. Joining a client to the Windows domain means enabling Kerberos as default protocol for authentication. The domain is a trust boundary similar to a federation on local level. Consequently, it is also a drawback, as user accounts and services need to have a trust relationship to the Kerberos token server. While Kerberos is utilized for computers, it can be used for, e.\,g., intranet web applications. Many UNIX like operating systems include software for Kerberos authentication of users or services. The reference architecture for FIM is applied for Kerberos. The adaptation is described from top to bottom per layer, see Figure~\ref{fig:kerberos}.

\begin{description}
\item[External Layer:] The layer consists of users with a client application for authentication. Identities, roles, and permissions are left out for simplicity.
\item[Business Service Layer:] Authentication is provided to a client by tickets, making SSO available. Some sort of policies need to be followed. The key exchange is important for mutual trust.
\item[Business Layer:] Authentication engine describes the authentication process of Kerberos, which has quality of service parameters. The authentication is basically provided by the technical service layer, described below. As authentication is only possible within trust boundaries, trust needs to be established.
\item[Application Service Layer:] The main service provided to both, user and server, is authentication, i.\,e., Kerberos tickets. In order to do this, the end points need to be discovered.
\item[Application Layer:] Kerberos is typically integrated into software, like AD, which utilizes the protocols LDAP, DNS, and Kerberos. This can be combined with FIM, shown in the section beforehand.
\item[Technical Service Layer:] AD internally uses several different server applications running on Windows servers. Besides LDAP and DNS server, this is mainly the Key Distribution Center (KDC), consisting of Kerberos ticket-granting service (TGS) and Kerberos Authentication Server (KAS). The authentication server forwards the username to KDC, which issues a ticket-granting ticket, stamped and encrypted by TGS.
\end{description}

\section{Conclusion and Outlook}
\label{sec:conclusion}
Internet services and, thereby, identity management are more relevant than ever before. Different protocols were developed, while further extensions and revisions are in the making. The multitude of standards and solutions challenges users as well as system administrators to get an outline. The lack of interoperability further calls for a reference architecture to improve the current identity management landscape. In order to gain an overview and, therefore, an understanding of the services, we designed a reference service with the supporting toolkit of Enterprise Architecture using the framework ArchiMate. 

Based on a motivating scenario, we identified four core requirements and three research questions. These are not fulfilled by current approaches targeting identity management and modelling. Therefore, we introduced FIMSM. It consists of several layers, describing a multitude of aspects related to the term service. The reference architecture includes external layer, business service layer, business layer, application service layer, application layer, and the technical service layer. It distinguishes services from applications, providing a better overview. Our reference model FIMSM is compliant with and extends the I\&AM approach of ArchiMate itself. It integrates into the best practice solutions of ITIL, establishes the connection to TOGAF, and is conform to the Federated Mission Network, while using the standardized nomenclature. It thus also forms the link to the UML profile I\&AM and the ontologies for service offer, negotiation, and contract for federations. In order to discuss FIMSM, we applied the motivating scenario and analysed the reference architecture based on requirements and research questions. In comparison to current architectures, our model is more fine-grained and at the same time generic. It is the first universal model and conform with frameworks. With the application of FIMSM to centralized I\&AM, i.\,e., LDAP respectively Kerberos, we see that it can be adapted.

In future work, we plan to detail FIMSM for FIM with submodels and interfaces to business processes. This will help to establish federated processes in accordance to the components. We further aim to extend FIMSM to other identity management systems, e.\,g., for IoT and PAM, and models, e.\,g., user-centric and centralized identity management. This will assist to improve and extend the FIMSM with the goal to build a universal model for generic identity management. Last but not least, we intent to model further security related processes, mechanisms, and services.

%
%
%
 \bibliographystyle{splncs04}
\nobibliography{foo.bib}

\end{document}